\documentclass[preprint, 3p, review]{elsarticle}

\usepackage{lineno,hyperref,graphicx}
\modulolinenumbers[5]

\usepackage{amssymb, amsthm, amsmath, siunitx}
\usepackage{adjustbox}
\usepackage{slashed}

\newcommand{\UD}{{U(1)_{D}}}
\newcommand{\DF}[1]{{\xi_{(#1)}}}
\newcommand{\DFC}[1]{{\xi^{c}_{(#1)}}}

\journal{PLB}

\begin{document}

\begin{frontmatter}

%
%
%
%

\title{Tree level Majorana neutrino mass from Type-1 $\times$ Type-2 Seesaw mechanism with Dark Matter}

\author{Chi-Fong Wong\corref{coau}}
\cortext[coau]{Corresponding author}
\ead{cfwong.freeman@gmail.com}
\author{Yang Chen}
\address{Department of Mathematics, Faculty of Science and Technology, University of Macau, Macau SAR, China}

\begin{abstract}
We propose a type of hybrid Seesaw model that combines Type-1 and Type-2 Seesaw mechanism in multiplicative way to generate tree level Majorana neutrino mass and provides a Dark Matter candidate.
The model extends the Standard Model by extra gauge symmetry $\UD$ and hidden sector consisted of chiral fermions and additional scalar fields.
After spontaneous symmetry breaking, light neutrino masses are generated not only by exchange of the new heavy fermions as Type-1 Seesaw, but also by coupling to the naturally small induced vacuum expectation value of new heavy scalar as Type-2 Seesaw.
An unbroken residue of $\UD$ protects the lightest Dirac fermion required by anomaly cancellation in hidden sector from decaying, therefore giving rise to a Dark Matter candidate.
Due to strong enough Seesaw suppression from our hybridization, new physics scale can be as low as TeV in this model and discovering signal from LHC data is possible in near future.
\end{abstract}


\end{frontmatter}



\section{Introduction}
\label{sec:intro}

Neutrino oscillation indicates nonzero neutrino masses and mixings and that the Standard Model (SM) must be extended.
Minimal approaches generating Majorana neutrino mass at tree level such as the Type-1 \cite{Minkowski:1977sc, GellMann:1980vs, Yanagida:1979as, Mohapatra:1979ia}, Type-2 \cite{Konetschny:1977bn, Magg:1980ut, Schechter:1980gr, Cheng:1980qt}, and Type-3 \cite{Foot:1988aq} Seesaw models require new degrees of freedom (DOFs) that are too heavy to test in near future, although these models are well motivated by UV physics such as the Grand Unified Theory.
Hybridization of Seesaw mechanisms can provide stronger suppression on neutrino mass therefore lowering the mass scale of new physics (NP), for example,
by considering cancellation on neutrino masses between Type-1 and Type-2 contributions in ``Type-1 $+$ Type-2'' Seesaw models \cite{Akhmedov:2006de,Akhmedov:2006yp,Chao:2007mz},
or by doubling the Seesaw suppression with an extended sterile fermion sector in Inverse Seesaw models \cite{Mohapatra:1986bd}.

Particularly, Seesaw suppression of different types can be combined multiplicatively in a class of models \cite{Babu:2009aq,Kumericki:2012bh,McDonald:2013kca,Picek:2009is,Liao:2010cc,McDonald:2013hsa} where Type-3 Seesaw is generalized by replacing the fermion triplets and Higgs doublet in Yukawa coupling with higher dimensional representations.
The resulting neutrino masses are then suppressed not only by the intermediate fermions as in usual Type-3 models, but also by a naturally small induced vacuum expectation value (VEV) of the scalar multiplet as in Type-2 models.
We may refer this class of models ``Type-3 $\times$ Type-2'' models.
Beside technical success, UV origin of such hybridization will be curious to ask.

Another evidence of NP is the existence of Dark Matter (DM) and it is intriguing to understand both neutrino mass and DM with a common theory.
For example, in Type-1 Seesaw models, DM can be given by introducing a keV-scale sterile neutrino \cite{Boyarsky:2018tvu}, but it is not apparent the theoretical necessity of both neutrino mass and DM to each other.
In contrast, DM can be an inevitable consequence of neutrino mass due to specific representation of fields under the known gauge symmetry.
For example, in Minimal Dark Matter model \cite{Cirelli:2005uq}, neutral component of exotic multiplets introduced for neutrino mass generation can be accidentally stable thus can account for DM in the Universe.
Moreover, symmetry can be the common reason behind stability of DM and tininess of neutrino mass.
In some radiative models \cite{Krauss:2002px,Ma:2006km}, both DM decay and tree level neutrino mass are forbidden by a new global symmetry, rendering loop correction from dark particles the leading source of neutrino mass.
These approaches establish theoretical connection between both phenomena, however, they cannot provide further information about dark sector such as the number and couplings of the dark species.

In this work, we study a TeV-scale SM extension to address the origins of neutrino mass and DM by a hybrid ``Type-1 $\times$ Type-2'' Seesaw mechanism and an inevitable DM candidate, respectively, all originated from a set of new DOFs of which both number and representations are well motivated by symmetry.
Specifically, the SM is extended with a hidden sector consisted of a new gauge symmetry $\UD$, a set of SM-singlet chiral fermions, and additional scalar fields for symmetry breaking and communication to the SM sector.
Because of the chirality of fermions, anomaly cancellation must come into play.
The $\UD$ charge assignment $z_{1}, z_{2}, ..., z_{N}$ of the $N$ chiral fermions must satisfy equations $\sum_{i} z_{i} = 0$ and $\sum_{i} z^{3}_{i} = 0$, for cancellation of mixed gauge-gravitational anomaly \cite{Delbourgo:1972xb, Eguchi:1976db, AlvarezGaume:1983ig} and triangular gauge anomaly \cite{Adler:1969gk, Bell:1969ts, Bardeen:1969md}, respectively.
As a consequence, both number and couplings of hidden fermions are well determined.
Such approach of introducing a gauged chiral sector as the source of NP has advantage on predictive power and its analog to the SM \cite{deGouvea:2015pea,Berryman:2016rot,Wong:2020obo}.
In general, infinite number of solutions can be found satisfying the equations above.
Some of them have been reported in literature \cite{Wong:2020obo,Batra:2005rh,Sayre:2005yh,Nakayama:2011dj}.
In this work, the charge assignment of hidden fermions is given by following anomaly-free solution reported previously \cite{Wong:2020obo} \footnote{Also appears in \cite{deGouvea:2015pea}.}:
\begin{equation}\label{eq:chiral_fermion_charges}
    1, 2, 2, -3, -3, -3, 4.
\end{equation}
After the SSB of $\UD$ at TeV scale by the VEVs of two scalar singlets, the chiral fermions merge into two sets of mass eigenstates.
One set consisted of three Majorana fermions will be responsible for neutrino mass generation, another set consisted of two Dirac fermions acquires a residual symmetry thus by which the lightest Dirac fermion is stable and becomes the DM candidate.
Subsequently, when the Higgs doublet VEV is developed and the SM gauge symmetry is broken, the messenger scalar doublet that communicates between hidden sector and SM sector will acquire a naturally small VEV induced by its trilinear coupling with other scalars.
Consequently, light neutrino masses are generated by exchange of the TeV-scale Majorana fermions and Yukawa couplings proportional to the small induced VEV.
We refer such model the ``Type-1 $\times$ Type-2'' Seesaw model.
Because of the stronger suppression effect, the induced VEV can be as large as $\mathcal{O}(10 \text{GeV})$, with all constraints satisfied.
It gives rise to interesting signature in colliders observable in the $3000 \text{fb}^{-1}$ 14 TeV LHC data.

We discuss the basic structure and features of this model in sec. \ref{sec:model}, followed by explanation on neutrino mass generation in sec. \ref{sec:mass}, and other phenomenologies including DM and LHC signature in sec. \ref{sec:pheno}.
We conclude in sec. \ref{sec:concl}.

\section{Model}
\label{sec:model}

The SM is extended by a new gauge symmetry $\UD$ and a hidden sector that contains a set of chiral fermions $\DF{z_{i}}$ singlet under the SM gauge group $SU(3)_{C} \times SU(2)_{L} \times U(1)_{Y}$ while carrying respective $\UD$ charge $z_{i}$ given by eq. (\ref{eq:chiral_fermion_charges}).
Without lost of generality, all chiral fermions are assumed right-handed.
The model also considers an extended scalar sector.
Beside the SM-like Higgs doublet $\Phi$, we also introduce a messenger scalar doublet $\eta$ that connects the hidden sector and the SM lepton sector, and two scalar singlets $S_{3}$ and $S_{6}$ responsible for spontaneous breaking of $\UD$ and generates masses to all fermionic DOFs\footnote{Instead, introducing two scalar singlets $S_{1}$ and $S_{2}$ carrying one and two units of $\UD$ charge respectively can play the same role on symmetry breaking and mass generation, while it leads to only one Majorana fermion and three Dirac fermions after symmetry breaking.}.
The scalar fields are parametrized by
\begin{equation}
    \Phi =
    \begin{pmatrix}
    \phi^{+} \\
    \frac{v + \phi^{0}_{R} + i \phi^{0}_{I}}{\sqrt{2}}
    \end{pmatrix}
    ,
    \quad
    \eta =
    \begin{pmatrix}
    \eta^{+} \\
    \frac{w + \eta^{0}_{R} + i \eta^{0}_{I}}{\sqrt{2}}
    \end{pmatrix}
    ,
    \quad
    S_{3} =
    \frac{u_{3} + s_{3R} + i s_{3I}}{\sqrt{2}}
    ,
    \quad
    S_{6} =
    \frac{u_{6} + s_{6R} + i s_{6I}}{\sqrt{2}}
\end{equation}
The particle content is shown in Table \ref{table:particle}.
\begin{table}
\begin{center}
    \begin{tabular}{c|ccccc|cccc}
    \hline
    & $L_{1,2,3}$ & $\DF{1}$ & $\DF{2}_{1,2}$ & $\DF{-3}_{1,2,3}$ & $\DF{4}$ & $\Phi$ & $\eta$ & $S_{3}$ & $S_{6}$
    \\\hline
    $SU(2)_{L}$ & \textbf{2} & \textbf{1} & \textbf{1} & \textbf{1} & \textbf{1} & \textbf{2} & \textbf{2} & \textbf{1} & \textbf{1}
    \\
    $U(1)_{Y}$ & -1/2 & 0 & 0 & 0 & 0 & 1/2 & 1/2 & 0 & 0
    \\
    $U(1)_{D}$ & 0 & 1 & 2 & -3 & 4 & 0 & -3 & 3 & 6
    \\\hline
    \end{tabular}
\end{center}
\caption{\label{table:particle}
Particle content of the model.
$L_{i}$ and $\Phi$ the SM lepton doublet and Higgs scalar doublet respectively.}
\end{table}

The mass spectrum of the fermions is determined by following Yukawa sector
\begin{equation}
    - \mathcal{L}
    \supset g_{ij} \overline{\DFC{-3}_{i}} \DF{-3}_{j} S_{6}
    + h_{l} \overline{\DFC{2}_{l}} \DF{1} S_{3}^{\ast}
    + k_{l} \overline{\DFC{2}_{l}} \DF{4} S_{6}^{\ast}
    + h.c.,
\end{equation}
where $i, j = 1,2,3$ and $l = 1, 2$.
The coupling matrix $g_{ij}$ is symmetric and can be defined real and diagonal $\text{diag}(g_{i})$.
At low energy, the nonzero vacuum expectation values (VEVs) $\langle S_{3} \rangle =  u_{3} / \sqrt{2}$ and $\langle S_{6} \rangle =  u_{6} / \sqrt{2}$ lead to three Majorana fermion mass eigenstates $N_{i} = \DF{-3}_{i} + \DFC{-3}_{i}$ with masses $m_{N_{i}} = \sqrt{2} g_{i} u_{6}$.
Moreover, the chiral fermions $\DF{1}$, $\DF{2}_{1,2}$, and $\DF{4}$ are mixed via
\begin{equation}\label{eq:dm_mass_matrix}
    - \mathcal{L}
    \supset
    \frac{1}{\sqrt{2}}
    \begin{pmatrix}
        \overline{\DFC{2}_{1}} &
        \overline{\DFC{2}_{2}}
    \end{pmatrix}
    \begin{pmatrix}
        h_{1} u_{3} & k_{1} u_{6} \\
        h_{2} u_{3} & k_{2} u_{6}
    \end{pmatrix}
    \begin{pmatrix}
        \DF{1} \\
        \DF{4}
    \end{pmatrix}
    + h.c.
\end{equation}
to generate two Dirac fermions $\Psi_{1}$ and $\Psi_{2}$.
The arbitrary complex 2-by-2 mass matrix in eq. (\ref{eq:dm_mass_matrix}) can be diagonalized with a bi-unitary transformation consisted of unitary matrices $U_{L}$ and $U_{R}$, giving rise $(\Psi_{1}, \Psi_{2})^T = U_{L}^{\dagger} (\DFC{2}_{1}, \DFC{2}_{2})^T + U_{R}^{\dagger} (\DF{1}, \DF{4})^T$.

These Dirac fermions experience flavor off-diagonal interaction stemmed from two sources.
One from Yukawa interactions with the CP-even scalar bosons $s_{3R}$ and $s_{6R}$, and CP-odd scalar bosons $s_{3I}$ and $s_{6I}$ in
\begin{equation}\label{eq:dm_yukawa_coupling}
    - \sqrt{2} \mathcal{L}
    \supset
    \begin{pmatrix}
        \overline{\Psi_{1L}} &
        \overline{\Psi_{2L}}
    \end{pmatrix}
    \left[
        U_{L}^{\dagger}
        \begin{pmatrix}
            h_{1} & 0 \\
            h_{2} & 0
        \end{pmatrix}
        U_{R}
        (s_{3R} + i s_{3I})
        +
        U_{L}^{\dagger}
        \begin{pmatrix}
            0 & k_{1} \\
            0 & k_{2}
        \end{pmatrix}
        U_{R}
        (s_{6R} + i s_{6I})
    \right]
    \begin{pmatrix}
        \Psi_{1R} \\
        \Psi_{2R}
    \end{pmatrix}
    + h.c.
\end{equation}
Another source of flavor off-diagonal interaction is from gauge interaction between the right-handed component of $\Psi_{i}$ and the new gauge boson $X_{\mu}$:
\begin{equation}\label{eq:dm_gauge_coupling}
    \mathcal{L}
    \supset
    g_{D}
    X_{\mu}
    \begin{pmatrix}
        \overline{\Psi_{1R}} &
        \overline{\Psi_{2R}}
    \end{pmatrix}
    \gamma^{\mu}
    \left[
        U_{R}^{\dagger}
        \begin{pmatrix}
            1 & 0 \\
            0 & 4
        \end{pmatrix}
        U_{R}
    \right]
    \begin{pmatrix}
        \Psi_{1R} \\
        \Psi_{2R}
    \end{pmatrix}
\end{equation}
where $g_{D}$ and $X_{\mu}$ are the coupling and gauge field corresponding to $\UD$, respectively.
In general, the Yukawa coupling matrices in eq. (\ref{eq:dm_yukawa_coupling}) and the charge matrix $U_{R}^{\dagger} \text{diag}(1, 4) U_{R}$ in eq. (\ref{eq:dm_gauge_coupling}) are not diagonal.
Therefore, the heavier Dirac fermion $\Psi_{2}$ can decay into the lighter one $\Psi_{1}$ plus an on/off-shell scalar $s_{3R}$, $s_{3I}$, $s_{6R}$, $s_{6I}$, or vector boson $X^{\mu}$.

The most general renormalizable scalar potential is given by \footnote{A similar but not identical scalar potential has been studied in \cite{Nomura:2017jxb}.}
\begin{equation}
    \begin{aligned}
    \mathcal{V}
    = & - \mu_{1}^{2} | \Phi |^{2} + \mu_{2}^{2} | \eta |^{2} - \mu_{3}^{2} | S_{3} |^{2} - \mu_{4}^{2} | S_{6} |^{2}
    \\
    & + \lambda_{1} | \Phi |^{4} + \lambda_{2} | \eta |^{4} + \lambda_{3} | S_{3} |^{4} + \lambda_{4} | S_{6} |^{4}
    \\
    & + \lambda_{12} | \Phi |^{2} | \eta |^{2} + \lambda_{12}^{\prime} | \Phi^{\dagger} \eta |^{2}
     + \lambda_{13} | \Phi |^{2} | S_{3} |^{2} + \lambda_{14} | \Phi |^{2} | S_{6} |^{2}
    \\
    & + \lambda_{23} | \eta |^{2} | S_{3} |^{2} + \lambda_{24} | \eta |^{2} | S_{6} |^{2}
    + \lambda_{34} | S_{3} |^{2} | S_{6} |^{2}
    \\
    & + \kappa_{1} \Phi^{\dagger} \, \eta \, S_{3} + \kappa_{2} S_{3} \, S_{3} \, S_{6}^{\ast} + \lambda \Phi^{\dagger} \, \eta \, S_{3}^{\ast} \, S_{6} + h.c.
    \end{aligned}
\end{equation}
The non-Hermitian operators in the last line break the global $[U(1)]^{4}$ symmetry in the scalar potential to $U(1)$, thus one may expect the parameters $\kappa_{1}$, $\kappa_{2}$, and $\lambda$ are naturally small.
Moreover, only two out of three complex phases of these parameters can be absorbed by the scalar fields, so one of them is complex in principle, providing possibility of explicit CP violation.

This scalar potential provides two interesting scenarios about the VEVs of the scalar fields.
In the first scenario, only $- \mu_{1}^{2}$ and $- \mu_{3}^{2}$ among all four quadratic terms are negative, giving rise to nonzero $v$ and $u_{3}$.
However, the coupling $\mathcal{L} \supset \kappa_{2} S_{3} \, S_{3} \, S_{6}^{\ast}$ induces the VEV $u_{6} \sim \kappa_{2} u_{3}^{2} / \mu_{4}^{2}$ that could be Type-2 ``Seesaw-ed'' if $\kappa_{2}$ is regarded a small parameter.
Furthermore, together with $v$, coupling $\mathcal{L} \supset \kappa_{1} \Phi^{\dagger} \, \eta \, S_{3}$ induces the fourth VEV $w$ that is also ``Seesaw-ed'' as $w \sim \kappa_{1} v u_{3} / \mu_{2}^{2}$.
This scenario gives rise to hierarchy $u_{3} \gg u_{6}$ and $u_{3}, v \gg w$.
An apparent consequence is that the first column in mass matrix of $\Psi_{i}$'s in eq. (\ref{eq:dm_mass_matrix}) will be significantly larger than the second column, resulting in large mass splitting between $\Psi_{1}$ and $\Psi_{2}$.
More explicitly, one can have $m_{\Psi_{2}} \gg m_{\Psi_{1}}, m_{N_{i}}$.

Another scenario assumes all four quadratic couplings are negative except $\mu_{2}^{2}$, giving rise to nonzero $v$, $u_{3}$, and $u_{6}$.
The operator $\mathcal{L} \supset \kappa_{1} \Phi^{\dagger} \, \eta \, S_{3} + \lambda \Phi^{\dagger} \, \eta \, S_{3}^{\ast} \, S_{6}$ then induces the Seesaw VEV $w \sim (\kappa_{1} v u_{3} + \lambda v u_{3} u_{6}) / \mu_{2}^{2}$.
If we assume $u_{3} \sim u_{6}$, we should have all $m_{\Psi_{1}}$, $m_{\Psi_{2}}$, and $m_{N_{i}}$ at the similar mass scale.
In this work, we adopt the second scenario assuming both $u_{3}$ and $u_{6}$ are generated by the negative quadratic terms and are at TeV scale.
The only hierarchically small VEV is $w$, and it will play important role in neutrino mass generation.

The more complete expression of the VEVs can be calculated by the first-order derivative to the scalar potential.
To simplify the result, we can assume $(\kappa_{1}, \kappa_{2}, \lambda u_{3}, \lambda u_{6}) \ll (u_{3}, u_{6})$, therefore the leading order (expanded in respect to $v$) VEVs are
\begin{subequations}
    \begin{align}
    u_{3} \simeq & \ \sqrt{\frac{4 \lambda_{4} \mu_{3}^{2} - 2 \lambda_{34} \mu_{4}^{2}}{4 \lambda_{3} \lambda_{4} - \lambda_{34}^{2}}}
    \\
    u_{6} \simeq & \ \sqrt{\frac{4 \lambda_{3} \mu_{4}^{2} - 2 \lambda_{34} \mu_{3}^{2}}{4 \lambda_{3} \lambda_{4} - \lambda_{34}^{2}}}
    \\
    v \simeq & \ \sqrt{\frac{2 \mu_{1}^{2} - \lambda_{13} u_{3}^{2} - \lambda_{14} u_{6}^{2}}{2 \lambda_{1}}}
    \\
    w \simeq & \ \frac{- (\sqrt{2} \kappa_{1} + \lambda u_{6}) v u_{3}}{2 \mu_{2}^{2} + (\lambda_{12} + \lambda_{12}^{\prime}) v^{2} + \lambda_{23} u_{3}^{2} + \lambda_{24} u_{6}^{2}} \label{eq:w_formula}
    \end{align}
\end{subequations}

This model therefore suggests an almost complete breaking of the gauge $\UD$ symmetry.
The second scalar doublet $\eta$, scalar singlets $S_{3}$ and $S_{6}$, and the heavy Majorana fermions $N_{i}$ are not charged under any residual symmetry thus all of them mix with SM DOFs of the same quantum numbers.
The charged scalars $\phi^{\pm}$ and $\eta^{\pm}$ are mixed into the physical charged scalar $H^{\pm}$ and the charged Goldstone bosons $G^{\pm}$ being absorbed by $W^{\pm}$.
The CP-odd neutral scalars $\phi^{0}_{I}$, $\eta^{0}_{I}$, $s_{3I}$ and $s_{6I}$ are mixed into two neutral Goldstone bosons $G_{1}$ and $G_{2}$ being absorbed by the two neutral gauge bosons, leaving two physical pseudoscalar bosons $A_{1}$ and $A_{2}$.
The CP-even neutral scalars $\phi^{0}_{R}$, $\eta^{0}_{R}$, $s_{3R}$ and $s_{6R}$ are mixed into four physical scalar bosons $H_{i}$ with $i = 1,2,3,4$.
The lightest state $H_{1}$ can be interpreted as the scalar boson of mass 125 GeV discovered at the LHC \cite{Tanabashi:2018oca}.
Finally, the three Majorana fermions $N_{i}$ mix with the three active neutrinos, giving rise three more mass eigenstates, as discussed in next section.
The only exception is the Dirac fermions $\Psi_{1,2}$.
They acquire accidentally conversed global $U(1)$ quantum number, forming a dark sector where the lightest state ($\Psi_{1}$) is stable and will play the role of Weakly Interactive Mass Particle (WIMP) DM.

\section{Neutrino mass}
\label{sec:mass}

\begin{figure}
    \centering
    \includegraphics[width=6.6cm]{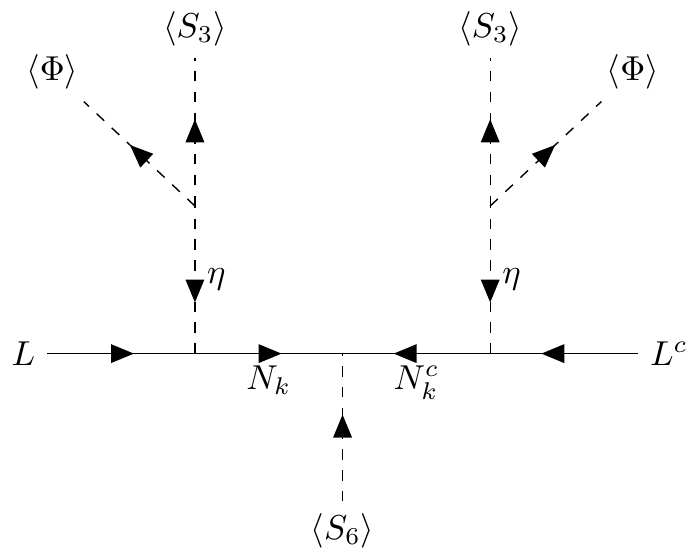}
\caption{\label{fig:neutrino_mass}
Neutrino mass generated at tree level.
Only contribution from $\Phi^{\dagger} \, \eta \, S_{3}$ operator is shown.}
\end{figure}

The messenger scalar doublet $\eta$ connects the SM lepton doublet $L$ and the chiral hidden fermions $\DF{3}_{i}$ via Yukawa couplings:
\begin{equation}
    \mathcal{L}
    = - y_{ij} \overline{L_{i}} \tilde{\eta} \DF{3}_{j} + h.c.
    \xrightarrow{SSB}
    - \frac{y_{ij} w}{\sqrt{2}} \overline{\nu_{iL}} N_{j} + h.c.
\end{equation}
With the Majorana mass terms $\mathcal{L} \supset - (m_{N_{i}}/2) \overline{N_{i}} N_{i}$, the active neutrinos obtain the Type-1 Seesaw Majorana neutrino masses after SSB:
\begin{equation}\label{eq:neutrino_mass}
    m^{\nu}_{ij} = \sum_{k} \frac{y^{\ast}_{ik} y^{\ast}_{jk} w^{2}}{m_{N_{k}}}
\end{equation}
Recall that $w$ is induced by operators $\kappa_{1} \Phi^{\dagger} \, \eta \, S_{3}$ and $\lambda \Phi^{\dagger} \, \eta \, S_{3}^{\ast} \, S_{6}$ in scalar potential, the tree level neutrino mass generation can be given by Feynman diagram shown in fig. \ref{fig:neutrino_mass}.
This process corresponds to a dim-8 operator $(1/\Lambda^{4}) \overline{L} L^{c} \Phi^{\dagger} \Phi S_{3} S_{3} S_{6}^{\ast}$.
There is also contribution from two loop-level that can be constructed by connecting the two $S_{3}$ legs with the $S_{6}$ leg in fig. \ref{fig:neutrino_mass} through operator $\kappa_{2} S_{3} \, S_{3} \, S_{6}^{\ast}$.
Although such diagram corresponds to dim-5 Weinberg operator, loop suppression, propagators, and $\kappa_{2}$ suggests it a subleading contribution.

It is worth to compare the new physics scale suggested by this model and that by the conventional Seesaw mechanisms.
In Type-1 and Type-3 Seesaw, neutrino masses are determined by propagator of heavy right-handed neutrinos of mass scale $M$ and Yukawa coupling proportional to the VEV $v$ of Higgs doublet;
in Type-2 Seesaw, neutrino masses are determined directly by Yukawa coupling proportional to the Seesaw-suppressed VEV $\mu_\text{tri} v^{2} / M^{2}$ of a heavy scalar triplet of mass scale $M$, where $\mu_\text{tri}$ denotes the trilinear coupling in scalar potential;
in this model, neutrino masses are determined not only by propagator of the heavy right-handed neutrinos and Yukawa coupling proportional to VEV $w$, but also the Seesaw-suppression subjected by $w \sim (\kappa_{1} + \lambda u_{6}) v u_{3} / \mu_{2}^{2}$.
To make its analogy to Type-2 Seesaw models more apparent, we denote $w \sim \mu_\text{tri} v / M$ in our comparison.
If we assume all relevant Yukawa coupling about $\mathcal{O}(1)$, $v \sim \mathcal{O}(100 \si{GeV})$, and neutrino mass about $\mathcal{O}(0.01 \si{eV})$, the new physics scale $M$ in each type of Seesaw mechanism can be estimated:
\begin{subequations}
    \begin{align}
    \text{Type-1/3:}
    \quad &
    M \sim 10^{15} \si{GeV}
    \\
    \text{Type-2:}
    \quad &
    M \sim 10^{15} \si{GeV} \ \frac{\mu_\text{tri}}{M}
    \\
    \text{Type-1 $\times$ Type-2 (this model):}
    \quad &
    M \sim 10^{15} \si{GeV} \ \left(\frac{\mu_\text{tri}}{M}\right)^{2}
    \end{align}
\end{subequations}
Assuming a small trilinear coupling $\mu_\text{tri} \sim \mathcal{O}(1 \si{MeV})$, we have $M \sim 10^{6} \si{GeV}$ for Type-2 Seesaw, while $M \sim 1 \si{TeV}$ for Type-1 $\times$ Type-2 Seesaw (this model) due to the double suppression from twice appearance of the Type-2 Seesaw suppression (on $w$).
This estimation shows that this model is more testable than the conventional Seesaw models in near future experiments.

Therefore, this model shares part of phenomenologies with the Type-1 Seesaw models, such as neutrino oscillation \cite{Brdar:2019iem}, non-unitarity in leptonic flavor mixings \cite{Antusch:2006vwa}, and collider signatures \cite{Ng:2015hba}, except the vanilla leptogenesis due to too light new physics scale \cite{Davidson:2002qv, Giudice:2003jh}.

\section{Phenomenology}
\label{sec:pheno}

\subsection{Rho parameter}

In Two Higgs Doublet Model, the condensate of the second scalar doublet does not affect $\rho$ parameter at tree level if it shares the same EW charges with the SM Higgs doublet.
In this model, however, the present of the gauge $\UD$ under which $\eta$ is charged brings about mass mixing between the two massive neutral gauge vector bosons $Z$ and $Z^{\prime}$, even though ignoring the kinetic mixing $\mathcal{L} \supset (-\epsilon/2) X^{\mu \nu} B_{\mu \nu}$ where $B_{\mu \nu}$ the field strength of hypercharge gauge field.
The $\rho = 1 + \delta \rho = m_{W} / m_{Z} \cos \theta_\text{weak}$ is then deviated from unity at tree level as
\begin{equation}\label{eq:delta_rho}
    \delta \rho_\text{tree}
    = \frac{(w/v)^{4}}{1 + (w/v)^{2}} \left[ \left(\frac{u_{3} z_{S_{3}}}{v z_{\eta}}\right)^{2}
    + \left(\frac{u_{6} z_{S_{6}}}{v z_{\eta}}\right)^{2}
    - \frac{g_{1}^{2} + g_{2}^{2}}{4 g_{D}^{2} z_{\eta}^{2}} \right]^{-1}
\end{equation}
where $z_{a}$ is the $\UD$ charge of particle $a = \eta, S_{3}, S_{6}$. 
The global fit of EW Precision Tests in PDG \cite{Tanabashi:2018oca} gives rise $\rho_\text{exp, fit} = 1.00038 \pm 0.00020$.
Within $1 \sigma$ uncertainty, $w$ can take values as large as $\mathcal{O}(10 \si{GeV})$ if $v = 246$ GeV, $u_{3} = u_{6} = 1$ TeV, and $g_{D} = 0.2$.


\subsection{LFV processes}

The charged scalar bosons $H^{\pm}$ mediate Lepton Flavor Violating (LFV) decays of charged leptons, e.g., $\mu \to e \gamma$, at one loop-level via Yukawa coupling $y_{ij} \overline{\ell_{i}} N_{j} \eta^{-}$.
Relevant calculation is the same as in the scotogenic models \cite{Ma:2006km,Vicente:2014wga}.
In the large $w$-regime where one may consider $w = 10$ GeV, eq. (\ref{eq:neutrino_mass}) suggests a generic Yukawa coupling $y \sim \mathcal{O}(10^{-5})$, that leads to tiny $\text{BR}(\mu^{+} \to e^{+} \gamma) \sim 10^{-28}$ far away from the MEG bound \cite{Baldini:2020okg}.
LFV processes such as $\mu \to e \gamma$ in this model becomes relevant to near future experiments only in small $w$-regime where assuming $w = 10 - 100$ MeV, leading to $y \sim 10^{-1} - 10^{-2}$ thus $\text{BR}(\mu^{+} \to e^{+} \gamma) \sim 10^{-13} - 10^{-16}$.

\subsection{DM}

Dirac fermions $\Psi_{1}$ and $\Psi_{2}$ are charged under an unbroken residual $U(1)$ symmetry after SSB, therefore the lightest state (i.e., $\Psi_{1}$) is stable and plays the role of WIMP DM.
DM is thermally produced in the early universe through its couplings with gauge boson $X$ and scalar bosons $S_{3}$ and $S_{6}$ that further mix with the SM DOFs.
However, these portals cannot play the dominant role on DM relic density because the direct DM search experiments have put stringent constraints.
Following analysis in \cite{Escudero:2016gzx,Arcadi:2017kky}, the coupling constants for Higgs-portal operator $\overline{\Psi_{1}} \Psi_{1} H_{1}$ and for Z-portal operator $\overline{\Psi_{1}} \gamma^{\mu} \Psi_{1} X_{\mu}$ have upper limits about $10^{-3}$ and $10^{-5}$, respectively, if DM mass is relatively light (100 GeV).
In this model, the new scalar boson $s_{3R}$ mixes with $\phi^{0}_{R}$ through the 4-by-4 symmetric mass-squared matrix
\begin{equation}\label{eq:scalar_mass_matrix}
	\begin{pmatrix}
		\lambda_{1} v^2 &
		\frac{1}{2} u_3 (\sqrt{2} \kappa_{1} + \lambda u_{6}) &
		\lambda_{13} v u_{3} &
		\lambda_{14} v u_{6}
		\\
		\dots &
		\frac{-1}{2} \frac{v}{w} u_3 (\sqrt{2} \kappa_{1} + \lambda u_{6}) &
		\frac{1}{2} v (\sqrt{2} \kappa_{1} + \lambda u_{6}) &
		\frac{1}{2} \lambda v u_{3} &
		\\
		\dots &
		\dots &
		2 \lambda_{3} u_{3}^2 &
		u_3 (\sqrt{2} \kappa_{2} + \lambda_{34} u_{6}) &
		\\
		\dots &
		\dots &
		\dots &
		2 \lambda_{4} u_{6}^{2} - \frac{1}{\sqrt{2}} \frac{\kappa_{2}}{u_{6}} u_{3}^{2}
	\end{pmatrix}
\end{equation}
in the basis of $(\phi^{0}_{R}, \eta^{0}_{R}, s_{3R}, s_{6R})$.
The DM candidate $\Psi_{1}$ interacts with Higgs-like bosons $H_{1}$ through $\mathcal{L} \supset a \overline{\Psi_{1}} \Psi_{1} s_{3R} \sim a \sin \theta \overline{\Psi_{1}} \Psi_{1} H_{1}$, where $a$ denotes the Yukawa coupling from the 1-1 component of the first non-diagonal coupling matrix shown in eq. (\ref{eq:dm_yukawa_coupling}) after the bi-unitary transformation $U_{L}$ and $U_{R}$; and $\theta$ the mixing angle between $\phi^{0}_{R}$ and $s_{3R}$ that can be estimated from eq. (\ref{eq:scalar_mass_matrix}) is about $\theta \sim (\lambda_{13} v) / (2 \lambda_{3} u_{3})$, thus $a \sin \theta \lesssim 10^{-3}$ is expected given $v = 246 \text{GeV}$ and $u_{3} = 1 \text{TeV}$ and appropriate values for $a$ and the $\lambda$'s.
Similar result can be obtained in case for $S_{6}$.
Moreover, the mixing angle between $Z$ and $Z^{\prime}$ is given by:
\begin{equation}
    \theta_{Z}
    = \frac{2 \sqrt{g_{1}^{2} + g_{2}^{2}} g_{D} z_{\eta} w^{2}}{4 g_{D}^{2} (w^{2} z_{\eta}^{2} + u_{3}^{2} z_{S_{3}}^{2} + u_{6}^{2} z_{S_{6}}^{2}) - (g_{1}^{2} + g_{2}^{2}) (v^{2}+w^{2})} + \mathcal{O}(w^{4})
\end{equation}
and it can be estimated $\theta_{Z} \sim (w/u_{3})^{2} \sim 10^{-4}$ as $w \simeq 10 \si{GeV}$ and $u_{3} \simeq 1 \si{TeV}$, giving rise to $g_{D} \theta_{Z} \lesssim \mathcal{O}(10^{-5})$ that satisfies the relevant bound. 
In this sense, the large $w$-regime of this model can be made compatible with Higgs- and Z-portal constraints from direct DM searches if DM mass is 100 GeV.
The constraints are even looser with heavier DM candidate.

DM relic abundance is then dominantly determined by its annihilation into the unstable new particles in this model.
For example, $\Psi_{1} \overline{\Psi_{1}} \to N_{i} N_{i}$ mediated by $s_{6R}$ in s-channel can give rise to thermal averaged cross-section
\begin{equation}
    \langle \sigma v \rangle (\Psi_{1} \overline{\Psi_{1}} \to N_{1} N_{1})
    \sim \frac{\lambda_{S}^{4} m_{\Psi_{1}}^{2}}{m_{S}^{4}}
\end{equation}
estimated by dimensional analysis,
where $m_{S}$ the mass of the mediator scalar boson $s_{6R}$ and $\lambda_{S}$ the coupling of the scalar to the fermions.
A large enough cross-section $\langle \sigma v \rangle \simeq 1 \text{pb}$ demanded by observational result $\Omega_\text{DM} h^{2} \simeq 0.1$ from Planck \cite{Aghanim:2018eyx} through relation $\Omega_{\Psi_{1}} h^{2} = 0.1 \text{pb} / \langle \sigma v \rangle$ will be satisfied if $\lambda_{S} \simeq 0.5$, $m_{S} = 1 \text{TeV}$, and DM mass $m_{\Psi_{1}} = 200 \text{GeV}$.

The same annihilation process $\Psi_{1} \overline{\Psi_{1}} \to N_{i} N_{i}$ also happen in the present-day universe in zero-momentum limit and provides possible signal to indirect DM searches.
If considering RHN decays to SM species through only active-sterile mixing, DM mass of about 200 GeV satisfies constraints from gamma-ray \cite{Campos:2017odj, Folgado:2018qlv}, see also that including anti-proton data \cite{Batell:2017rol}.


\subsection{LHC signal}

\begin{figure}
    \center
    \includegraphics[width=7.6cm]{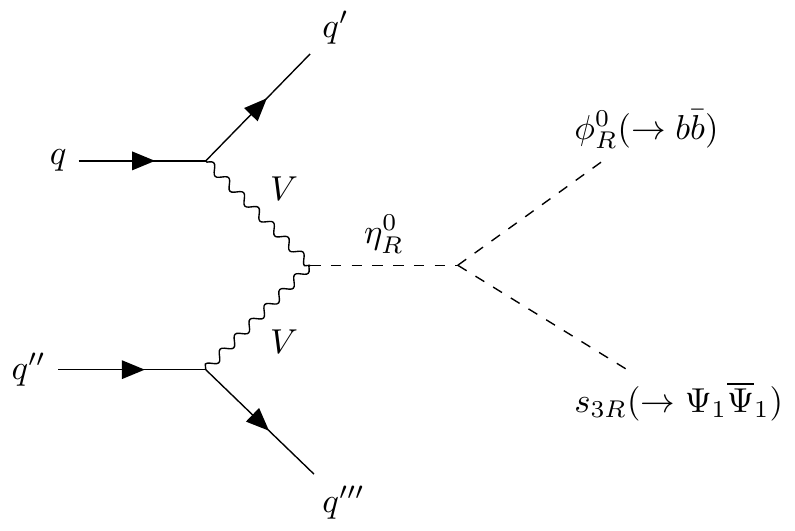}
\caption{\label{fig:lhc_single_production}
Single production of the messenger scalar boson $\eta_{R}^{0}$ and its subsequent decay into boosted $b \bar{b}$ pair and large missing transverse energy, where $V$ stands for either W or Z boson of the SM.}
\end{figure}

Given the merit of low energy scale for new physics, new particles introduced in this model are more hopeful to produce and detect in collider experiments.
Especially, the crucial trilinear coupling $\kappa_{1}$ that helps to determine the neutrino mass scale can be directly probed.
Because $\eta$ does not carry any residual symmetry from the hidden sector, it can be singly produced in the LHC.
For example, $\eta_{R}^{0}$ can be singly produced via Vector-Boson-Fusion(VBF)-like process in which a pair of gauge bosons, either $W^{\pm}$ or $Z^{0}$, fuse into $\eta^{0}_{R}$.
Trilinear coupling $\kappa_{1} \Phi^{\dagger} \eta S_{3}$ then mediates the subsequent decay of $\eta_{R}^{0} \to \phi_{R}^{0} s_{3R}$.
The $s_{3R}$ is assumed heavy and decays dominantly into a pair of DM particle $\Psi_{1} \overline{\Psi}_{1}$, giving rise to large missing transverse energy ($\slashed{E}_{T}$, MET), while the $\phi_{R}^{0} \simeq H_{1}$ is then highly boosted and decays into a pair of bottom quark and antiquark.
The overall signature seen in detectors will be consisted of a boosted ``Higgs-jet'' identified by the constituent $b$ and $\bar{b}$ jets, plus large MET, plus two quark jets.
The process is depicted in Fig. \ref{fig:lhc_single_production}.

The cross-section of such process can be roughly estimated.
Assuming all intermediate states are produced on-shell and their widths are negligible (Narrow Width Approximation), in flavor basis, we have
\begin{equation}
    \begin{aligned}
        \sigma(p p \to b \bar{b} q \bar{q} \slashed{E}_{T})
        \simeq & \ 0.051 \text{fb}
        \left(\frac{\sigma(p p \to q \bar{q} \eta_{R}^{0})}{106 \text{fb} \left(\frac{10}{246}\right)^{2}}\right)
        \left(\frac{\text{BR}(\eta_{R}^{0} \to \phi_{R}^{0} s_{3R})}{0.5}\right)
        \\
        & \ \times
        \left(\frac{\text{BR}(\phi_{R}^{0} \to b \bar{b})}{0.58}\right)
        \left(\frac{\text{BR}(s_{3R} \to \Psi_{1} \overline{\Psi}_{1})}{1}\right)
    \end{aligned}
\end{equation}
The VBF-like production cross-section $\sigma(p p \to q \bar{q} \eta_{R}^{0})$ can be estimated from the similar process of the SM Higgs boson \cite{LHCHiggsCrossSectionWorkingGroup:2011wcg}, scaled by $(w/v)^2$.
In the large $w$-regime where $w = 10$ GeV, we can have $\sigma(p p \to q \bar{q} \eta_{R}^{0}) \sim 0.18$ fb if $m_{\eta_{R}^{0}} = 1$ TeV is assumed.
We assume the branching ratios $\text{BR}(\eta_{R}^{0} \to \phi_{R}^{0} s_{3R}) \sim \mathcal{O}(0.1)$ and $\text{BR}(s_{3R} \to \Psi_{1} \overline{\Psi}_{1}) \sim 1$ as they are controlled by model parameters.
The branching ratio $\text{BR}(\phi_{R}^{0} \to b \bar{b}) \simeq \text{BR}(H_{1} \to b \bar{b})$ is known about 0.58.
In high luminosity run of the LHC, this process could generate about 153 events if integrated luminosity is assumed $3000 \text{fb}^{-1}$, and such level of event samples may be possible for a discovery in near future.
This process is informative for model parameters.
The invariant mass of the MET can be found just the mass of $s_{3R}$;
the invariant mass of the $b \bar{b}$-pair, on the other hand, is just the mass of $\phi_{R}^{0}$ which is assumed known as 125 GeV.
The invariant mass of the $b \bar{b}$-pair and the MET together is just the mass of the $\eta_{R}^{0}$.
Moreover, the total cross-section depends on $\kappa_{1}$, $w$, and $\text{BR}(s_{3R} \to \Psi_{1} \overline{\Psi}_{1})$.
Once discovery, these parameters can be further determined and/or constrained.

Similar but not identical event signature consisted of a boosted Higgs decaying into $b \bar{b}$ pair plus large MET has been searched for in the LHC and reported by ATLAS \cite{ATLAS:2021shl} and CMS \cite{CMS:2019ykj}.
In these searches, pair of DM particles are generated in final state associated with a boosted Higgs boson.
Ref. \cite{ATLAS:2021shl} has given rise model-independent upper limit on cross-section between 0.05 - 3.26 fb, depending on MET.
However, the process we have discussed contains two extra quark jets appear in final state, that is not considered in the experimental searches above, so the quoted limit is not directly applicable in this case.



\section{Conclusion}
\label{sec:concl}

To go beyond the minimal Seesaw models for additional features such as testability and Dark Matter while still being UV-motivated, we have extended the SM by a new gauge symmetry $\UD$, a set of anomaly-free SM singlet chiral fermions, and an enlarged scalar sector.
After SSB of $\UD$ and EW symmetry, the new chiral fermions compose two sets of massive fermions that play important role in neutrino physics and cosmology.
One set of fermions consisted of three Majorana fermions play role as the right-handed neutrinos and are responsible for generating tree level neutrino masses via the Type-1 Seesaw mechanism;
another set of fermions consisted of two Dirac fermions carry a residual unbroken symmetry of the $\UD$ and provides a DM candidate.
In contrast to put-in-by-hand, both sets of fermions rely on existence of each other intrinsically as required by anomaly cancellation.

Unlike conventional Type-1 Seesaw models, the mass scale of the right-handed neutrinos can naturally be around TeV scale thanks to the additional suppression originated from Type-2-like Seesaw in scalar sector gives rise to a tiny induced VEV that lowers the Dirac mass terms between left- and right-handed neutrinos.
Therefore, such hybrid Seesaw mechanism is referred to Type-1 $\times$ Type-2 Seesaw model since the suppression from both sources are combined multiplicatively.
The model suggests a lower mass scale of new physics, and allows a relatively large value of the induced VEV $w$.
With $w \simeq 10$ GeV, constraints from rho parameter, LFV decays of charged leptons, DM relic density and direct/indirect searches are all satisfied, while an interesting collider signature is possible for discovery in the upcoming High-Luminosity run of the LHC at CERN.
Decay of a singly produced scalar boson of $\eta$ from VBF-like process in colliders can give rise to final state consisted of a highly boosted $b \bar{b}$ jet, large missing transverse energy, and two quark jets.
Once measured, such process can reveal information including the couplings and masses of the relevant new particles.
With a small $w$, instead, the interactions between SM sector and the new sector such as $Z$-$Z^{\prime}$ mixing and new particle production in colliders become more difficult, while LFV decays of charged leptons will be closer to experimental sensitivity in near future.






\section*{Acknowledgments}
This work was supported by The Science and Technology Development Fund (FDCT) of Macau SAR [grant number 0079/2020/A2].
Feynman diagrams in this paper are drawn by Tikz-Feynman package \cite{Ellis:2016jkw}.


\bibliography{main}

\begin{thebibliography}{10}
\expandafter\ifx\csname url\endcsname\relax
  \def\url#1{\texttt{#1}}\fi
\expandafter\ifx\csname urlprefix\endcsname\relax\def\urlprefix{URL }\fi
\expandafter\ifx\csname href\endcsname\relax
  \def\href#1#2{#2} \def\path#1{#1}\fi

\bibitem{Minkowski:1977sc}
P.~Minkowski, {$\mu \to e\gamma$ at a Rate of One Out of $10^{9}$ Muon
  Decays?}, Phys. Lett. B 67 (1977) 421--428.
\newblock \href {https://doi.org/10.1016/0370-2693(77)90435-X}
  {\path{doi:10.1016/0370-2693(77)90435-X}}.

\bibitem{GellMann:1980vs}
M.~Gell-Mann, P.~Ramond, R.~Slansky, {Complex Spinors and Unified Theories},
  Conf. Proc. C 790927 (1979) 315--321.
\newblock \href {http://arxiv.org/abs/1306.4669} {\path{arXiv:1306.4669}}.

\bibitem{Yanagida:1979as}
T.~Yanagida, {Horizontal gauge symmetry and masses of neutrinos}, Conf. Proc. C
  7902131 (1979) 95--99.

\bibitem{Mohapatra:1979ia}
R.~N. Mohapatra, G.~Senjanovic, {Neutrino Mass and Spontaneous Parity
  Nonconservation}, Phys. Rev. Lett. 44 (1980) 912.
\newblock \href {https://doi.org/10.1103/PhysRevLett.44.912}
  {\path{doi:10.1103/PhysRevLett.44.912}}.

\bibitem{Konetschny:1977bn}
W.~Konetschny, W.~Kummer, {Nonconservation of Total Lepton Number with Scalar
  Bosons}, Phys. Lett. B 70 (1977) 433--435.
\newblock \href {https://doi.org/10.1016/0370-2693(77)90407-5}
  {\path{doi:10.1016/0370-2693(77)90407-5}}.

\bibitem{Magg:1980ut}
M.~Magg, C.~Wetterich, {Neutrino Mass Problem and Gauge Hierarchy}, Phys. Lett.
  B 94 (1980) 61--64.
\newblock \href {https://doi.org/10.1016/0370-2693(80)90825-4}
  {\path{doi:10.1016/0370-2693(80)90825-4}}.

\bibitem{Schechter:1980gr}
J.~Schechter, J.~W.~F. Valle, {Neutrino Masses in SU(2) x U(1) Theories}, Phys.
  Rev. D 22 (1980) 2227.
\newblock \href {https://doi.org/10.1103/PhysRevD.22.2227}
  {\path{doi:10.1103/PhysRevD.22.2227}}.

\bibitem{Cheng:1980qt}
T.~P. Cheng, L.-F. Li, {Neutrino Masses, Mixings and Oscillations in SU(2) x
  U(1) Models of Electroweak Interactions}, Phys. Rev. D 22 (1980) 2860.
\newblock \href {https://doi.org/10.1103/PhysRevD.22.2860}
  {\path{doi:10.1103/PhysRevD.22.2860}}.

\bibitem{Foot:1988aq}
R.~Foot, H.~Lew, X.~G. He, G.~C. Joshi, {Seesaw Neutrino Masses Induced by a
  Triplet of Leptons}, Z. Phys. C 44 (1989) 441.
\newblock \href {https://doi.org/10.1007/BF01415558}
  {\path{doi:10.1007/BF01415558}}.

\bibitem{Akhmedov:2006de}
E.~K. Akhmedov, M.~Frigerio, {Interplay of type I and type II seesaw
  contributions to neutrino mass}, JHEP 01 (2007) 043.
\newblock \href {http://arxiv.org/abs/hep-ph/0609046}
  {\path{arXiv:hep-ph/0609046}}, \href
  {https://doi.org/10.1088/1126-6708/2007/01/043}
  {\path{doi:10.1088/1126-6708/2007/01/043}}.

\bibitem{Akhmedov:2006yp}
E.~K. Akhmedov, M.~Blennow, T.~Hallgren, T.~Konstandin, T.~Ohlsson, {Stability
  and leptogenesis in the left-right symmetric seesaw mechanism}, JHEP 04
  (2007) 022.
\newblock \href {http://arxiv.org/abs/hep-ph/0612194}
  {\path{arXiv:hep-ph/0612194}}, \href
  {https://doi.org/10.1088/1126-6708/2007/04/022}
  {\path{doi:10.1088/1126-6708/2007/04/022}}.

\bibitem{Chao:2007mz}
W.~Chao, S.~Luo, Z.-z. Xing, S.~Zhou, {A Compromise between Neutrino Masses and
  Collider Signatures in the Type-II Seesaw Model}, Phys. Rev. D 77 (2008)
  016001.
\newblock \href {http://arxiv.org/abs/0709.1069} {\path{arXiv:0709.1069}},
  \href {https://doi.org/10.1103/PhysRevD.77.016001}
  {\path{doi:10.1103/PhysRevD.77.016001}}.

\bibitem{Mohapatra:1986bd}
R.~N. Mohapatra, J.~W.~F. Valle, {Neutrino Mass and Baryon Number
  Nonconservation in Superstring Models}, Phys. Rev. D 34 (1986) 1642.
\newblock \href {https://doi.org/10.1103/PhysRevD.34.1642}
  {\path{doi:10.1103/PhysRevD.34.1642}}.

\bibitem{Babu:2009aq}
K.~S. Babu, S.~Nandi, Z.~Tavartkiladze, {New Mechanism for Neutrino Mass
  Generation and Triply Charged Higgs Bosons at the LHC}, Phys. Rev. D 80
  (2009) 071702.
\newblock \href {http://arxiv.org/abs/0905.2710} {\path{arXiv:0905.2710}},
  \href {https://doi.org/10.1103/PhysRevD.80.071702}
  {\path{doi:10.1103/PhysRevD.80.071702}}.

\bibitem{Kumericki:2012bh}
K.~Kumericki, I.~Picek, B.~Radovcic, {TeV-scale Seesaw with Quintuplet
  Fermions}, Phys. Rev. D 86 (2012) 013006.
\newblock \href {http://arxiv.org/abs/1204.6599} {\path{arXiv:1204.6599}},
  \href {https://doi.org/10.1103/PhysRevD.86.013006}
  {\path{doi:10.1103/PhysRevD.86.013006}}.

\bibitem{McDonald:2013kca}
K.~L. McDonald, {Minimal Tree-Level Seesaws with a Heavy Intermediate Fermion},
  JHEP 07 (2013) 020.
\newblock \href {http://arxiv.org/abs/1303.4573} {\path{arXiv:1303.4573}},
  \href {https://doi.org/10.1007/JHEP07(2013)020}
  {\path{doi:10.1007/JHEP07(2013)020}}.

\bibitem{Picek:2009is}
I.~Picek, B.~Radovcic, {Novel TeV-scale seesaw mechanism with Dirac mediators},
  Phys. Lett. B 687 (2010) 338--341.
\newblock \href {http://arxiv.org/abs/0911.1374} {\path{arXiv:0911.1374}},
  \href {https://doi.org/10.1016/j.physletb.2010.03.062}
  {\path{doi:10.1016/j.physletb.2010.03.062}}.

\bibitem{Liao:2010cc}
Y.~Liao, {Cascade Seesaw for Tiny Neutrino Mass}, JHEP 06 (2011) 098.
\newblock \href {http://arxiv.org/abs/1011.3633} {\path{arXiv:1011.3633}},
  \href {https://doi.org/10.1007/JHEP06(2011)098}
  {\path{doi:10.1007/JHEP06(2011)098}}.

\bibitem{McDonald:2013hsa}
K.~L. McDonald, {Probing Exotic Fermions from a Seesaw/Radiative Model at the
  LHC}, JHEP 11 (2013) 131.
\newblock \href {http://arxiv.org/abs/1310.0609} {\path{arXiv:1310.0609}},
  \href {https://doi.org/10.1007/JHEP11(2013)131}
  {\path{doi:10.1007/JHEP11(2013)131}}.

\bibitem{Boyarsky:2018tvu}
A.~Boyarsky, M.~Drewes, T.~Lasserre, S.~Mertens, O.~Ruchayskiy, {Sterile
  neutrino Dark Matter}, Prog. Part. Nucl. Phys. 104 (2019) 1--45.
\newblock \href {http://arxiv.org/abs/1807.07938} {\path{arXiv:1807.07938}},
  \href {https://doi.org/10.1016/j.ppnp.2018.07.004}
  {\path{doi:10.1016/j.ppnp.2018.07.004}}.

\bibitem{Cirelli:2005uq}
M.~Cirelli, N.~Fornengo, A.~Strumia, {Minimal dark matter}, Nucl. Phys. B 753
  (2006) 178--194.
\newblock \href {http://arxiv.org/abs/hep-ph/0512090}
  {\path{arXiv:hep-ph/0512090}}, \href
  {https://doi.org/10.1016/j.nuclphysb.2006.07.012}
  {\path{doi:10.1016/j.nuclphysb.2006.07.012}}.

\bibitem{Krauss:2002px}
L.~M. Krauss, S.~Nasri, M.~Trodden, {A Model for neutrino masses and dark
  matter}, Phys. Rev. D 67 (2003) 085002.
\newblock \href {http://arxiv.org/abs/hep-ph/0210389}
  {\path{arXiv:hep-ph/0210389}}, \href
  {https://doi.org/10.1103/PhysRevD.67.085002}
  {\path{doi:10.1103/PhysRevD.67.085002}}.

\bibitem{Ma:2006km}
E.~Ma, {Verifiable radiative seesaw mechanism of neutrino mass and dark
  matter}, Phys. Rev. D 73 (2006) 077301.
\newblock \href {http://arxiv.org/abs/hep-ph/0601225}
  {\path{arXiv:hep-ph/0601225}}, \href
  {https://doi.org/10.1103/PhysRevD.73.077301}
  {\path{doi:10.1103/PhysRevD.73.077301}}.

\bibitem{Delbourgo:1972xb}
R.~Delbourgo, A.~Salam, {The gravitational correction to pcac}, Phys. Lett. B
  40 (1972) 381--382.
\newblock \href {https://doi.org/10.1016/0370-2693(72)90825-8}
  {\path{doi:10.1016/0370-2693(72)90825-8}}.

\bibitem{Eguchi:1976db}
T.~Eguchi, P.~G.~O. Freund, {Quantum Gravity and World Topology}, Phys. Rev.
  Lett. 37 (1976) 1251.
\newblock \href {https://doi.org/10.1103/PhysRevLett.37.1251}
  {\path{doi:10.1103/PhysRevLett.37.1251}}.

\bibitem{AlvarezGaume:1983ig}
L.~Alvarez-Gaume, E.~Witten, {Gravitational Anomalies}, Nucl. Phys. B 234
  (1984) 269.
\newblock \href {https://doi.org/10.1016/0550-3213(84)90066-X}
  {\path{doi:10.1016/0550-3213(84)90066-X}}.

\bibitem{Adler:1969gk}
S.~L. Adler, {Axial vector vertex in spinor electrodynamics}, Phys. Rev. 177
  (1969) 2426--2438.
\newblock \href {https://doi.org/10.1103/PhysRev.177.2426}
  {\path{doi:10.1103/PhysRev.177.2426}}.

\bibitem{Bell:1969ts}
J.~S. Bell, R.~Jackiw, {A PCAC puzzle: $\pi^0 \to \gamma \gamma$ in the
  $\sigma$ model}, Nuovo Cim. A 60 (1969) 47--61.
\newblock \href {https://doi.org/10.1007/BF02823296}
  {\path{doi:10.1007/BF02823296}}.

\bibitem{Bardeen:1969md}
W.~A. Bardeen, {Anomalous Ward identities in spinor field theories}, Phys. Rev.
  184 (1969) 1848--1857.
\newblock \href {https://doi.org/10.1103/PhysRev.184.1848}
  {\path{doi:10.1103/PhysRev.184.1848}}.

\bibitem{deGouvea:2015pea}
A.~de~Gouv\^ea, D.~Hern\'andez, {New Chiral Fermions, a New Gauge Interaction,
  Dirac Neutrinos, and Dark Matter}, JHEP 10 (2015) 046.
\newblock \href {http://arxiv.org/abs/1507.00916} {\path{arXiv:1507.00916}},
  \href {https://doi.org/10.1007/JHEP10(2015)046}
  {\path{doi:10.1007/JHEP10(2015)046}}.

\bibitem{Berryman:2016rot}
J.~M. Berryman, A.~de~Gouv\^ea, D.~Hern\'andez, K.~J. Kelly, {Imperfect mirror
  copies of the standard model}, Phys. Rev. D 94~(3) (2016) 035009.
\newblock \href {http://arxiv.org/abs/1605.03610} {\path{arXiv:1605.03610}},
  \href {https://doi.org/10.1103/PhysRevD.94.035009}
  {\path{doi:10.1103/PhysRevD.94.035009}}.

\bibitem{Wong:2020obo}
C.-F. Wong, {Anomaly-free chiral $U(1)_D$ and its scotogenic implication},
  Phys. Dark Univ. 32 (2021) 100818.
\newblock \href {http://arxiv.org/abs/2008.08573} {\path{arXiv:2008.08573}},
  \href {https://doi.org/10.1016/j.dark.2021.100818}
  {\path{doi:10.1016/j.dark.2021.100818}}.

\bibitem{Batra:2005rh}
P.~Batra, B.~A. Dobrescu, D.~Spivak, {Anomaly-free sets of fermions}, J. Math.
  Phys. 47 (2006) 082301.
\newblock \href {http://arxiv.org/abs/hep-ph/0510181}
  {\path{arXiv:hep-ph/0510181}}, \href {https://doi.org/10.1063/1.2222081}
  {\path{doi:10.1063/1.2222081}}.

\bibitem{Sayre:2005yh}
J.~Sayre, S.~Wiesenfeldt, S.~Willenbrock, {Sterile neutrinos and global
  symmetries}, Phys. Rev. D 72 (2005) 015001.
\newblock \href {http://arxiv.org/abs/hep-ph/0504198}
  {\path{arXiv:hep-ph/0504198}}, \href
  {https://doi.org/10.1103/PhysRevD.72.015001}
  {\path{doi:10.1103/PhysRevD.72.015001}}.

\bibitem{Nakayama:2011dj}
K.~Nakayama, F.~Takahashi, T.~T. Yanagida, {Number-Theory Dark Matter}, Phys.
  Lett. B 699 (2011) 360--363.
\newblock \href {http://arxiv.org/abs/1102.4688} {\path{arXiv:1102.4688}},
  \href {https://doi.org/10.1016/j.physletb.2011.04.035}
  {\path{doi:10.1016/j.physletb.2011.04.035}}.

\bibitem{Nomura:2017jxb}
T.~Nomura, H.~Okada, {Neutrinophilic two Higgs doublet model with dark matter
  under an alternative $U(1)_{B-L}$ gauge symmetry}, Eur. Phys. J. C 78~(3)
  (2018) 189.
\newblock \href {http://arxiv.org/abs/1708.08737} {\path{arXiv:1708.08737}},
  \href {https://doi.org/10.1140/epjc/s10052-018-5667-6}
  {\path{doi:10.1140/epjc/s10052-018-5667-6}}.

\bibitem{Tanabashi:2018oca}
M.~Tanabashi, et~al., {Review of Particle Physics}, Phys. Rev. D 98~(3) (2018)
  030001.
\newblock \href {https://doi.org/10.1103/PhysRevD.98.030001}
  {\path{doi:10.1103/PhysRevD.98.030001}}.

\bibitem{Brdar:2019iem}
V.~Brdar, A.~J. Helmboldt, S.~Iwamoto, K.~Schmitz, {Type-I Seesaw as the Common
  Origin of Neutrino Mass, Baryon Asymmetry, and the Electroweak Scale}, Phys.
  Rev. D 100 (2019) 075029.
\newblock \href {http://arxiv.org/abs/1905.12634} {\path{arXiv:1905.12634}},
  \href {https://doi.org/10.1103/PhysRevD.100.075029}
  {\path{doi:10.1103/PhysRevD.100.075029}}.

\bibitem{Antusch:2006vwa}
S.~Antusch, C.~Biggio, E.~Fernandez-Martinez, M.~B. Gavela, J.~Lopez-Pavon,
  {Unitarity of the Leptonic Mixing Matrix}, JHEP 10 (2006) 084.
\newblock \href {http://arxiv.org/abs/hep-ph/0607020}
  {\path{arXiv:hep-ph/0607020}}, \href
  {https://doi.org/10.1088/1126-6708/2006/10/084}
  {\path{doi:10.1088/1126-6708/2006/10/084}}.

\bibitem{Ng:2015hba}
J.~N. Ng, A.~de~la Puente, B.~W.-P. Pan, {Search for Heavy Right-Handed
  Neutrinos at the LHC and Beyond in the Same-Sign Same-Flavor Leptons Final
  State}, JHEP 12 (2015) 172.
\newblock \href {http://arxiv.org/abs/1505.01934} {\path{arXiv:1505.01934}},
  \href {https://doi.org/10.1007/JHEP12(2015)172}
  {\path{doi:10.1007/JHEP12(2015)172}}.

\bibitem{Davidson:2002qv}
S.~Davidson, A.~Ibarra, {A Lower bound on the right-handed neutrino mass from
  leptogenesis}, Phys. Lett. B 535 (2002) 25--32.
\newblock \href {http://arxiv.org/abs/hep-ph/0202239}
  {\path{arXiv:hep-ph/0202239}}, \href
  {https://doi.org/10.1016/S0370-2693(02)01735-5}
  {\path{doi:10.1016/S0370-2693(02)01735-5}}.

\bibitem{Giudice:2003jh}
G.~F. Giudice, A.~Notari, M.~Raidal, A.~Riotto, A.~Strumia, {Towards a complete
  theory of thermal leptogenesis in the SM and MSSM}, Nucl. Phys. B 685 (2004)
  89--149.
\newblock \href {http://arxiv.org/abs/hep-ph/0310123}
  {\path{arXiv:hep-ph/0310123}}, \href
  {https://doi.org/10.1016/j.nuclphysb.2004.02.019}
  {\path{doi:10.1016/j.nuclphysb.2004.02.019}}.

\bibitem{Vicente:2014wga}
A.~Vicente, C.~E. Yaguna, {Probing the scotogenic model with lepton flavor
  violating processes}, JHEP 02 (2015) 144.
\newblock \href {http://arxiv.org/abs/1412.2545} {\path{arXiv:1412.2545}},
  \href {https://doi.org/10.1007/JHEP02(2015)144}
  {\path{doi:10.1007/JHEP02(2015)144}}.

\bibitem{Baldini:2020okg}
A.~M. Baldini, et~al., {Search for lepton flavour violating muon decay mediated
  by a new light particle in the MEG experiment}, Eur. Phys. J. C 80~(9) (2020)
  858.
\newblock \href {http://arxiv.org/abs/2005.00339} {\path{arXiv:2005.00339}},
  \href {https://doi.org/10.1140/epjc/s10052-020-8364-1}
  {\path{doi:10.1140/epjc/s10052-020-8364-1}}.

\bibitem{Escudero:2016gzx}
M.~Escudero, A.~Berlin, D.~Hooper, M.-X. Lin, {Toward (Finally!) Ruling Out Z
  and Higgs Mediated Dark Matter Models}, JCAP 12 (2016) 029.
\newblock \href {http://arxiv.org/abs/1609.09079} {\path{arXiv:1609.09079}},
  \href {https://doi.org/10.1088/1475-7516/2016/12/029}
  {\path{doi:10.1088/1475-7516/2016/12/029}}.

\bibitem{Arcadi:2017kky}
G.~Arcadi, M.~Dutra, P.~Ghosh, M.~Lindner, Y.~Mambrini, M.~Pierre, S.~Profumo,
  F.~S. Queiroz, {The waning of the WIMP? A review of models, searches, and
  constraints}, Eur. Phys. J. C 78~(3) (2018) 203.
\newblock \href {http://arxiv.org/abs/1703.07364} {\path{arXiv:1703.07364}},
  \href {https://doi.org/10.1140/epjc/s10052-018-5662-y}
  {\path{doi:10.1140/epjc/s10052-018-5662-y}}.

\bibitem{Aghanim:2018eyx}
N.~Aghanim, et~al., {Planck 2018 results. VI. Cosmological parameters}, Astron.
  Astrophys. 641 (2020) A6, [Erratum: Astron.Astrophys. 652, C4 (2021)].
\newblock \href {http://arxiv.org/abs/1807.06209} {\path{arXiv:1807.06209}},
  \href {https://doi.org/10.1051/0004-6361/201833910}
  {\path{doi:10.1051/0004-6361/201833910}}.

\bibitem{Campos:2017odj}
M.~D. Campos, F.~S. Queiroz, C.~E. Yaguna, C.~Weniger, {Search for right-handed
  neutrinos from dark matter annihilation with gamma-rays}, JCAP 07 (2017) 016.
\newblock \href {http://arxiv.org/abs/1702.06145} {\path{arXiv:1702.06145}},
  \href {https://doi.org/10.1088/1475-7516/2017/07/016}
  {\path{doi:10.1088/1475-7516/2017/07/016}}.

\bibitem{Folgado:2018qlv}
M.~G. Folgado, G.~A. G\'omez-Vargas, N.~Rius, R.~Ruiz De~Austri, {Probing the
  sterile neutrino portal to Dark Matter with $\gamma$ rays}, JCAP 08 (2018)
  002.
\newblock \href {http://arxiv.org/abs/1803.08934} {\path{arXiv:1803.08934}},
  \href {https://doi.org/10.1088/1475-7516/2018/08/002}
  {\path{doi:10.1088/1475-7516/2018/08/002}}.

\bibitem{Batell:2017rol}
B.~Batell, T.~Han, B.~Shams Es~Haghi, {Indirect Detection of Neutrino Portal
  Dark Matter}, Phys. Rev. D 97~(9) (2018) 095020.
\newblock \href {http://arxiv.org/abs/1704.08708} {\path{arXiv:1704.08708}},
  \href {https://doi.org/10.1103/PhysRevD.97.095020}
  {\path{doi:10.1103/PhysRevD.97.095020}}.

\bibitem{LHCHiggsCrossSectionWorkingGroup:2011wcg}
S.~Dittmaier, et~al., {Handbook of LHC Higgs Cross Sections: 1. Inclusive
  Observables} (1 2011).
\newblock \href {http://arxiv.org/abs/1101.0593} {\path{arXiv:1101.0593}},
  \href {https://doi.org/10.5170/CERN-2011-002}
  {\path{doi:10.5170/CERN-2011-002}}.

\bibitem{ATLAS:2021shl}
G.~Aad, et~al., {Search for dark matter produced in association with a Standard
  Model Higgs boson decaying into b-quarks using the full Run 2 dataset from
  the ATLAS detector}, JHEP 11 (2021) 209.
\newblock \href {http://arxiv.org/abs/2108.13391} {\path{arXiv:2108.13391}},
  \href {https://doi.org/10.1007/JHEP11(2021)209}
  {\path{doi:10.1007/JHEP11(2021)209}}.

\bibitem{CMS:2019ykj}
A.~M. Sirunyan, et~al., {Search for dark matter particles produced in
  association with a Higgs boson in proton-proton collisions at $
  \sqrt{\mathrm{s}} $ = 13 TeV}, JHEP 03 (2020) 025.
\newblock \href {http://arxiv.org/abs/1908.01713} {\path{arXiv:1908.01713}},
  \href {https://doi.org/10.1007/JHEP03(2020)025}
  {\path{doi:10.1007/JHEP03(2020)025}}.

\bibitem{Ellis:2016jkw}
J.~Ellis, {TikZ-Feynman: Feynman diagrams with TikZ}, Comput. Phys. Commun. 210
  (2017) 103--123.
\newblock \href {http://arxiv.org/abs/1601.05437} {\path{arXiv:1601.05437}},
  \href {https://doi.org/10.1016/j.cpc.2016.08.019}
  {\path{doi:10.1016/j.cpc.2016.08.019}}.

\end{thebibliography}

\end{document}